\documentclass[11pt,a4paper]{article}
\usepackage{amsmath,amsfonts,amssymb,diagrams}
\usepackage{graphicx}

\newtheorem{theorem}{Theorem}[section]

\newenvironment{definition}[1][Definition]{\begin{trivlist}
\item[\hskip \labelsep {\bfseries #1}]}{\end{trivlist}}
\newenvironment{example}[1][Example]{\begin{trivlist}
\item[\hskip \labelsep {\bfseries #1}]}{\end{trivlist}}
\newenvironment{remark}[1][Remark]{\begin{trivlist}
\item[\hskip \labelsep {\bfseries #1}]}{\end{trivlist}}

\newcommand{\qed}{\nobreak \ifvmode \relax \else
      \ifdim\lastskip<1.5em \hskip-\lastskip
      \hskip1.5em plus0em minus0.5em \fi \nobreak
      \vrule height0.75em width0.5em depth0.25em\fi}

\begin{document}
\title{Dialectica Fuzzy Petri Nets }
\author{Valeria de Paiva \and Apostolos Syropoulos}
\date{January 2011}
\maketitle
\begin{abstract}
Brown and Gurr~\cite{brown90,brown95} have introduced a model of Petri Nets that is based on 
de~Paiva's Dialectica categories. This model was refined in an unpublished technical 
report~\cite{brown91}, where Petri nets  with multiplicities, instead of {\em elementary} nets 
(i.e., nets with multiplicities zero and one only) were considered. In this note we expand this modelling 
to deal with {\em fuzzy} petri nets. The basic idea is to use as the dualizing object in the Dialectica
categories construction, the unit interval that has all the properties of a {\em lineale} 
structure~\cite{syropoulos06}.
\end{abstract}
\section{The category $\mathsf{Dial}_{\mathrm{I}}(\mathbf{Set})$} 
The Dialectica categories construction~\cite{depaiva06} can be  instantiated using any lineale  
and the basic category $\mathbf{Set}$.  As discussed in~\cite{syropoulos06}, the unit interval, 
since it is a Heyting algebra, has all the properties of a lineale structure. In fact, one can 
prove that the quintuple $(\mathrm{I}, \le, \wedge, 1, \Rightarrow)$, where $\mathrm{I}$ is the 
unit interval, $a\wedge b=\min\{a,b\}$, and  $a\Rightarrow b=\bigvee\{c: c\wedge a\le b\}$ 
($a\vee b=\max\{a,b\}$),  is a lineale.

Let $U$ and $X$ be nonempty sets. A binary fuzzy relation $R$ in $U$ and $X$ is a fuzzy subset 
of $U\times X$, or $U\times X\to {\mathrm{I}}$. The relation $R(u,x)$ is interpreted as the degree of membership of the ordered pair $(u,x)$ in $R$. Let us now define a category of fuzzy relations.
 
The category $\mathsf{Dial}_{\mathrm{I}}(\mathbf{Set})$ has  as objects triples $A = (U, X, \alpha)$, 
where $U , X$ are sets and $\alpha$ is a map $U\times X\to\mathrm{I}$. Thus, each object
is literally a fuzzy relation.  A map from $A = (U, X, \alpha)$ to $B = (V, Y, \beta)$  
is a pair of  $\mathbf{Set}$ maps $(f,g)$,  $f:U\to V$, $g:Y\to X$ such that
\begin{displaymath}
\alpha(u, g(y)) \leq \beta(f(u),y),
\end{displaymath}
or in pictorial form:
\begin{diagram}
U\times Y                         & \rTo^{\mathrm{id}_{U}\times g}     & U\times X\\
\dTo^{f\times\mathrm{id}_{Y}}     & \ge                                & \dTo_{\alpha}\\
V\times Y                         & \rTo_{\beta}                       & \mathrm{I}
\end{diagram}
Assume that $(f,g)$ and $(f',g')$ are the following arrows:
\begin{displaymath}
(U,X,\alpha)\overset{(f,g)}{\underset{}{\longrightarrow}}
(V,Y,\beta)\overset{(f',g')}{\underset{}{\longrightarrow}}(W,Z,\gamma).
\end{displaymath}
Then $(f,g)\circ(f',g')=(f\circ f',g'\circ g)$ such that
\begin{displaymath}
\alpha\Bigl(u,\bigl( g'\circ g\bigr)(z)\Bigr) \le \gamma\Bigl(\bigl(f\circ f'\bigr)(u),z\Bigr).
\end{displaymath}

Tensor products  and the internal-hom in $\mathsf{Dial}_{\mathrm{I}}(\mathbf{Set})$ are given as  in the original Dialectica construction.
Given objects $A = (U, X, \alpha)$ and  $B = (V, Y, \beta)$, the tensor product  $A\otimes B$ is
$(U\times V, X^V\times Y^U, \alpha\times\beta)$ and the linear function-space or internal-hom is 
given by $A\to B= (V^U\times Y^X, U\times X, \alpha\to \beta)$. With this structure we obtain:
\begin{theorem}
The category ${\sf Dial}_{{\mathrm{I}}}({\bf Sets})$ is a  monoidal closed category.
\end{theorem}
Products and coproducts are given by $A\times B= (U\times V, X+ Y, \emptyset)$ and 
$A\oplus B= (U+ V, X\times Y, \emptyset)$.

Let $A = (U, X, \alpha)$ be an object of $\mathsf{Dial}_{\mathrm{I}}(\mathbf{Set})$, where $X$ is 
a frame. Then we can view  $A$ as a {\em fuzzy topological system}, that is, the fuzzy counterpart 
of Vickers's~\cite{vickers90} {\em topological systems}. A topological system is a triple
$(X,\models,A)$, where $A$ is a frame whose elements are called opens and $X$ is a set whose
elements are called points. Also, $\models$ is a subset of $X\times A$, and when $x\models a$, 
we say that $x$ {\em satisfies} $a$. In addition,
\begin{itemize}
\item If $S$ is a finite subset of $A$, then
\begin{displaymath}
x\models \bigwedge S \Longleftrightarrow x\models a\;\text{for all $a\in S$}.
\end{displaymath}
\item If $S$ is any subset of $A$, then 
\begin{displaymath}
x\models\bigvee S \Longleftrightarrow x\models a\;\text{for some $a\in S$}.
\end{displaymath}
\end{itemize}
Given two topological systems $(X,A)$ and $(Y,B)$, a function $f:X\rightarrow Y$ and a frame
homomorphism $\phi:B\rightarrow A$ are a {\em continuous} map from the first to the second
if $x\models \phi(b) \Leftrightarrow f(x)\models b$.

In order to fuzzify the structures just presented, we need to fuzzify the relation ``$\models$''
and nothing else: 
\begin{definition}
A {\em fuzzy topological system} is a triple $(X,\alpha,A)$, where $X$ is a set, $A$ is a frame
and $\alpha:X\times A\rightarrow\mathrm{I}$ a binary fuzzy relation such that
\begin{itemize}
\item If $S$ is a finite subset of $A$, then
\begin{displaymath}
\alpha(x, \bigwedge S)\ge0 \Longleftrightarrow \alpha(x,a)\ge0\;\text{for all $a\in S$}.
\end{displaymath}
\item If $S$ is any subset of $A$, then 
\begin{displaymath}
\alpha(x,\bigvee S)\ge0 \Longleftrightarrow \alpha(x,a)\ge0\;\text{for some $a\in S$}.
\end{displaymath}
\end{itemize} 
\end{definition}
\begin{remark}
Any topological system $(X,A)$ is a fuzzy topological system $(X,\iota,A)$, where
\begin{displaymath}
\iota(x,a)=\left\{\begin{array}{ll}
                   1, & \text{when $x\models a$}\\
                   0, & \text{otherwise}
                  \end{array}\right.
\end{displaymath}
\end{remark}
From the previous definition, one can easily deduce that
\begin{itemize}
\item $\alpha(x,\top)>0$ for all $x\in X$ and
\item $\alpha(x,\bot)=0$ for all $x\in X$.
\end{itemize}
\begin{example}
Vickers~\cite[p.~53]{vickers90} gives an interesting physical interpretation of topological systems.
In particular, he considers the set $X$ to be a set of programs that generate bit streams and
the opens to be assertions about bit streams. For exanple, if $x$ is a program that generates
the infinite bit stream 010101010101\ldots and ``\textbf{starts} 01010'' is an assertion that is
satified if a bit stream starts with the digits ``01010'', then this is expressed as follows: 
\begin{displaymath}
x\models \text{\textbf{starts} 01010}.
\end{displaymath}
Assume now that $x'$ is a program that produces bit streams that look like the following one
\begin{center}
\includegraphics[scale=1]{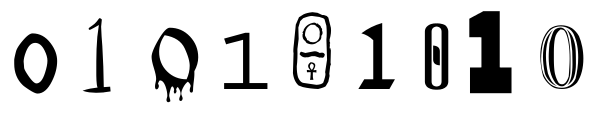}
\end{center}
The individuals bits are not identical to either ``1'' or ``0,'' but rather similar to these.
One can speculate that these bits are the result of some interaction of $x'$ with its environment
and this is the reason they are not identical. Then, we can say that $x'$ satisfies the assertion 
``\textbf{starts} 01010'' to some degree, since the elements that make up the stream produced by $x'$ 
are not identical, but rather similar.
\end{example}

The collection of objects of $\mathsf{Dial}_{\mathrm{I}}(\mathbf{Set})$ that are fuzzy topological
spaces and the arrows between them, form the category $\mathbf{FuzTopSpa}$, which is a
full subcategory of $\mathsf{Dial}_{\mathrm{I}}(\mathbf{Set})$. {\bfseries What about the category
of topological systems, is it a full subcategory of $\mathsf{Dial}_{\mathrm{I}}(\mathbf{Set})$?}

It is not difficult to map fuzzy topological systems to fuzzy topological spaces 
(see~\cite{ying-ming97}). The following definition shows how to map an open to fuzzy set:
\begin{definition}
Assume that $a\in A$, where $(X,\alpha,A)$ is a fuzzy topological space. Then the {\em extend}
of $a$ is a function whose graph is given below:
\begin{displaymath}
\Bigl\{ \bigl(x,\alpha(x,a) \bigr):x\in X\Bigr\}.
\end{displaymath}
\end{definition}
The collection of all fuzzy sets created by the extends of the members of $A$ corresponds to
a fuzzy topology on $X$. {\bfseries We need to show that this is indeed true.}

\section{The category $\mathsf{FNets}$} 
According to Winkel's insight~\cite{winskel88}, petri nets can be faithfully modeled as a set of 
{\em events} (usually denoted by $E$) and a set of {\em conditions} (written as $B$), related by two 
{\em multirelations},\footnote{A multirelation in $X$ and $Y$ is characterized by a
function $X\times Y\rightarrow\mathbb{N}$,  where $\mathbb{N}$ is the set of natural 
numbers (including zero).} corresponding to pre and post-conditions. Building from that, Brown, Gurr and 
de~Paiva~\cite{brown91} came up with the category ${\bf GNet}$, where morphisms of nets correspond 
to {\em simulations} and where the linear logic connectives correspond to net constructions. Below
we give  Winkel's definition~\cite{winskel87} of petri nets:
\begin{definition} 
A Petri Net is a quadruple $(E,B,\mathrm{pre},\mathrm{post})$, 
where $E$ and $B$ are  sets, and $\mathrm{pre}$ and $\mathrm{post}$ are multirelations
in $E \times B$.
\end{definition}
The following definition of  a fuzzy petri net differs significantly from ``standard'' definitions (e.g., 
see~\cite{chen90}), however it seems natural, if starting from Winskel's definition. Moreover, our 
objectives are quite different.
\begin{definition}
A {\em fuzzy} petri net is a quadruple $(E,B,\mathrm{pre},\mathrm{post})$, where $E$ and $B$ are 
sets, and $\mathrm{pre}$ and $\mathrm{post}$ are fuzzy relations in $E \times B$.
\end{definition}
Here, instead of an event $e$ being crisply related to a condition $b$, we now have a degree of 
relatedness between the event and the condition. We write $N$ for the Petri net 
$(E,B,\mathrm{pre},\mathrm{post})$, $N_0$ for the net $(E_0, B_0,\mathrm{pre}_0, \mathrm{post}_0)$, 
and so on.  We write $\mathbf{Petri}$ for the class of Petri nets and $\mathbf{FPetri}$ for the class 
of fuzzy Petri nets.
\begin{definition}
Consider the category $\mathbf{FNets}$ where objects are general fuzzy Petri nets, that is, elements of 
$\mathbf{FPetri}$, and a morphism from $(E, B,\mathrm{pre},\mathrm{post})$ to 
$(E', B',\mathrm{pre}',\mathrm{post}')$ is a pair of functions $(f, F)$ with $f:E\to E'$ and 
$F: B'\to B$ such that
\begin{center}
\begin{tabular}{ccc}
\begin{diagram}
E\times B'                         & \rTo^{\mathrm{id}_{E}\times F}     & E\times B\\
\dTo^{f\times\mathrm{id}_{B'}}     & \le                                & \dTo_{\mathrm{pre}}\\
E'\times B'                        & \rTo_{\mathrm{pre}'}               & \mathrm{I}
\end{diagram} & and &
\begin{diagram}
E\times B'                         & \rTo^{\mathrm{id}_{E}\times F}     & E\times B\\
\dTo^{f\times\mathrm{id}_{B'}}     & \ge                                & \dTo_{\mathrm{post}}\\
E'\times B'                        & \rTo_{\mathrm{post}'}              & \mathrm{I}
\end{diagram} 
\end{tabular}
\end{center}
that is, $\mathrm{pre}(e,F(b')\leq \mathrm{pre}'(f(e), b')$ and 
$\mathrm{post}'(e,F(b')\geq \mathrm{post}(f(e), b')$  $\forall e\in E$ and 
$\forall b'\in B'$.
\end{definition}
  
  \begin{theorem}
The category ${\sf  FNets}$ is a  monoidal closed category with products and coproducts, hence a model of modality-free linear logic.
\end{theorem}


\end{document}